\documentclass[twocolumn,showpacs,preprintnumbers,amsmath,amssymb]{revtex4}
\usepackage{graphicx}
\usepackage{dcolumn}
\usepackage{bm}

\begin{document}

\title{{\it Ab initio} studies of spin-spiral waves and exchange interactions  in 3{\it d} transition metal atomic chains}
\author{J. C. Tung$^1$ and G. Y. Guo$^{1,2}$\footnote{E-mail: gyguo@phys.ntu.edu.tw} }
\affiliation{$^1$Department of Physics and Center for Theoretical Sciences, National Taiwan University, Taipei
106, Taiwan\\$^2$Graduate Institute of Applied Physics, National Chengchi University, Taipei 116, Taiwan}
\date{\today}

\begin{abstract}
The total energy of the transverse spin-spiral wave as a function of the wave vector
for all 3$d$ transition metal atomic chains has been calculated
within  {\it ab initio} density functional theory with generalized
gradient approximation. It is predicted that at the equilibrium bond length, the V, Mn, and Fe
chains have a stable spin spiral structure, whilst the magnetic ground state of the Cr, Co and
Ni chains remains to be collinear.
Furthermore, all the exchange interaction parameters of the 3$d$ transition metal chains are evaluated
by using the calculated energy dispersion relations of the spin-spiral waves. Interestingly, it is found
that the magnetic couplings in the V, Mn and Cr chains are frustrated (i.e., the second near neighbor exchange
interaction is antiferromagnetic), and this leads to the formation of the stable spin-spiral structure
in these chains. The spin-wave stiffness constant of these 3$d$ metal chains is also evaluated and is found to be
smaller than its counterpart in bulk and monolayer systems.
The upper limit (in the order of 100 Kelvins) of the possible magnetic phase transition temperature
in these atomic chains is also estimated within the mean field approximation.
The electronic band structure of the spin-spiral structures have also been calculated.
It is hoped that the interesting findings here of the stable spin-spiral structure and frustrated magnetic
interaction in the 3$d$ transition metal chains would stimulate further theoretical
and experimental research in this field.

\end{abstract}

\pacs{71.70.Gm, 73.21.Hb, 75.10.-b, 75.30.Ds}

\maketitle

\section{Introduction}

Noncollinear magnetism, especially the spin-spiral structures, has received much attention in
recent decades, not only for
possible magnetism-based technological applications\cite{Kaj10,Lot04,Kim03} but also for
fundamental physics\cite{Bra09,Mor09,Lou08,Ferriani,Kat05,Lizarraga2,Tag01,Pajda2,Mry91,Tsunoda}.
In particular, it was recently reported that spin chirality in geometrically frustrated pyrochlore
compounds could generate magnetic monopoles\cite{Bra09,Mor09} and also large anomalous Hall effect\cite{Tag01}.
It was also proposed recently that the spin-spiral structure could be the main source of
the magnetoelectric effect observed recently in multiferroic oxides.\cite{Lot04,Kim03}
The spin-spiral structure, in which the magnetization rotates along a
certain direction in a bulk material, was observed two decades ago in neutron diffraction experiments on
fcc Fe and Fe$_{100_x}$Co$_x$ alloy precipitates in Cu.\cite{Tsunoda}
This experimental finding has since stimulated many {\it ab initio} studies of
the spin-spiral structures in bulk magnets\cite{Mry91,Uhl94,Kor96,Rosengaard,Halilov,Pajda,Marsman,Shallcross}.
Indeed, {\it ab initio} calculations\cite{Mry91,Halilov,Marsman}
corroborated that stable spin-spiral states exist in fcc Fe. Furthermore, {\it ab initio} total energy
calculations for the spin-spiral structures also helped to formulate an explanation of
the anomalous magnetovolume properties of the Invar alloys (the Invar effect)\cite{Uhl94}.


Noncollinear magnetism in low-dimensional systems has also been studied both theoretically and
experimentally in recent years.\cite{Ferriani,Mizuno,Nakamura,Crew}
For example, Mn monolayer on W(001) surface was recently investigated\cite{Ferriani} jointly
by spin-polarized scanning tunneling microscopy and also {\it ab initio} calculations,
and it was concluded that a spin-spiral structure along the (110) direction exists in
this monolayer system.
A stable spin-spiral structure with propagation vector ${\bf q}$ = (0,0,0.15)(2$\pi$/$a$)
was also predicted to exist in the unsupported freestanding Fe(110) monolayer with
the lattice constant of 3.16 \AA \cite{Mizuno,Nakamura}.
Co/CoPt bilayers were also found to support noncollinear spiral structures by Brillouin light
scattering\cite{Crew}. Interestingly, very recent {\it ab initio} calculations show that
in Mn chains on Ni(001), the magnetic structure could change from non-collinear to collinear ferrimagnetic,
depending on whether the number of the Mn atoms is even or odd.\cite{Lou08}


Stimulated by possible unusual magnetism in one-dimensional (1-D) systems, we have recently
carried out systematic {\it ab initio} studies of the collinear magnetic properties of
linear and zigzag atomic chains of all 3$d$\cite{Tung}, 4$d$ and 5$d$\cite{Tung2} transition metals.
Although the ideal infinite freestanding 3$d$ transition metal atomic chain is
unstable and cannot be prepared experimentally, short suspended monostrand metal nanowires and
atomic chains have been prepared in mechanical break junctions.\cite{ohn98,yan98,Rodrigues03} 
Furthermore, structurally stable Co atomic chains have been recently prepared on a vicinal surface of Pt(997)
surface\cite{Gam02} or inside nanotubes.\cite{Suh03} 
Therefore, we have also performed {\it ab initio} calculations for the 3$d$ transition metal
linear atomic chains on the Cu(001) surface\cite{tun11a,tun11b} in order to understand how the 
substrates would affect the magnetic properties of the nanowires. 

The purpose of the present work is to study possible spin-spiral structures in
all 3$d$ transition metal atomic chains by {\it ab initio} calculation of the total energy
of the spin-spiral state as a function of propagation wave vector ${\bf q}$.
Indeed, we find that the magnetic ground state in the V, Mn and Fe chains would be
a spin-spiral state.
Furthermore, we evaluate the exchange interaction parameters between the atoms and also spin-wave
stiffness constants of all the
atomic chains considered here from the calculated energy dispersion relations of the
spin-spiral waves. The obtained exchange interaction parameters allow us to understand
why the spin-spiral state is stable in the V, Mn and Fe chains but is not stable in
the Cr, Co and Ni chains. Finally, we also estimated the upper limits of the
magnetic phase transition temperature for all the atomic chains.


This paper is organized as follows. After a brief description of the computational details in Sec. II,
we present all the calculated energy dispersion relations of the spin-spiral waves
of the 3$d$ atomic chains in Sec. III. These results show that a stable spin-spiral state exists in
the V, Mn and Fe chains. Reported in Sec. IV are the obtained exchange interaction parameters
which enable us to understand the stability of the obtained magnetic ground state in
each atomic chain considered.
In Sec. V, we present the calculated  spin-wave stiffness constant and also the estimated
magnetic phase transition temperatures for the 3$d$ atomic chains.
Finally, the band structures of the spin-spiral state of the V and Mn chains are displayed in
Sec. VI, and a summary of this work is given in Sec. VII.

\section{Theory and Computational Method}

In the present first principles calculations, we use the accurate frozen-core full-potential
projector augmented-wave (PAW) method,~\cite{blo94} as implemented in the Vienna {\it ab initio}
simulation package (VASP) \cite{vasp1,vasp2}. The calculations are based on density functional theory
with the exchange and correlation effects being described by the generalized gradient approximation
(GGA)\cite{PW91}. A very large plane-wave cutoff energy of 500 eV is used. The shallow core 3$p$ electrons of the 3$d$
transition metals are treated as valence electrons. We adopt the standard supercell approach to model an
isolated atomic chain. The nearest wire and wire distance adopted here is 20~\AA.
We start with the theoretical equilibrium bond lengths for collinear magnetic states
from our previous study of 3$d$ TM nanowires.\cite{Tung}
However, in the fully unconstrained noncollinear magnetic calculations\cite{Hobbs} for the spin spiral structures,
we vary the bond length in order to study the bond length dependence of the stability of the spin spiral state.
The $\Gamma$-centered Monkhorst-Pack scheme with a $k$-mesh of $1\times1\times n$ ($ n = 100$) in the full
Brillouin zone (BZ), in conjunction with the Fermi-Dirac-smearing method with $\sigma = 0.02$ eV, is used
for the BZ integration.

We consider the transverse spin-spiral states where all the spins rotate in a plane perpendicular to
the spiral propagation vector {\bf q}. The total energies of the transverse spin-spirals as
a function of the magnitude of spin-spiral wave vector $q$ are calculated self-consistently
by using the generalized Bloch condition approach\cite{Herring,Sandratskii}. To study the exchange interactions,
we apply the frozen-magnon approach and obtain the exchange interaction parameters by a Fourier transformation
of the energy spectra of the spin-spiral waves.

\section{Stability of spin-spiral states}

\begin{table}
\caption{Calculated equilibrium bond length $d$,
ground state spin-spiral wave vector ($q$) and total energy [$E(q)$] [relative to that of the FM state ($q = 0$)]
as well as spin magnetic moment ($m_s$) at $q = 0$ in the 3$d$ transition metal chains.}
\begin{ruledtabular}
\begin{tabular}{ccccc}
          & $m_s$   & $d$    & $q$                   & $E(q)$  \\
          & ($\mu_B$/atom)   &  (\AA )   &  (2$\pi$/$d$)  & (meV/atom)\\ \hline
      V   & 1.47 &  2.05 &  0.25& -166.4\\
      Cr  & 4.18 &  2.32 &  0.50& -155.2\\
      Mn  & 4.43 &  2.40 &  0.33& -113.3\\
      Fe  & 3.30 &  2.25 &  0.10&  -15.5\\
      Co  & 2.18 &  2.15 &  0.00&  ---    \\
      Ni  & 1.14 &  2.18 &  0.00&  ---
\end{tabular}
\end{ruledtabular}
\label{table1}
\end{table}

\begin{figure}
\includegraphics[width=8cm]{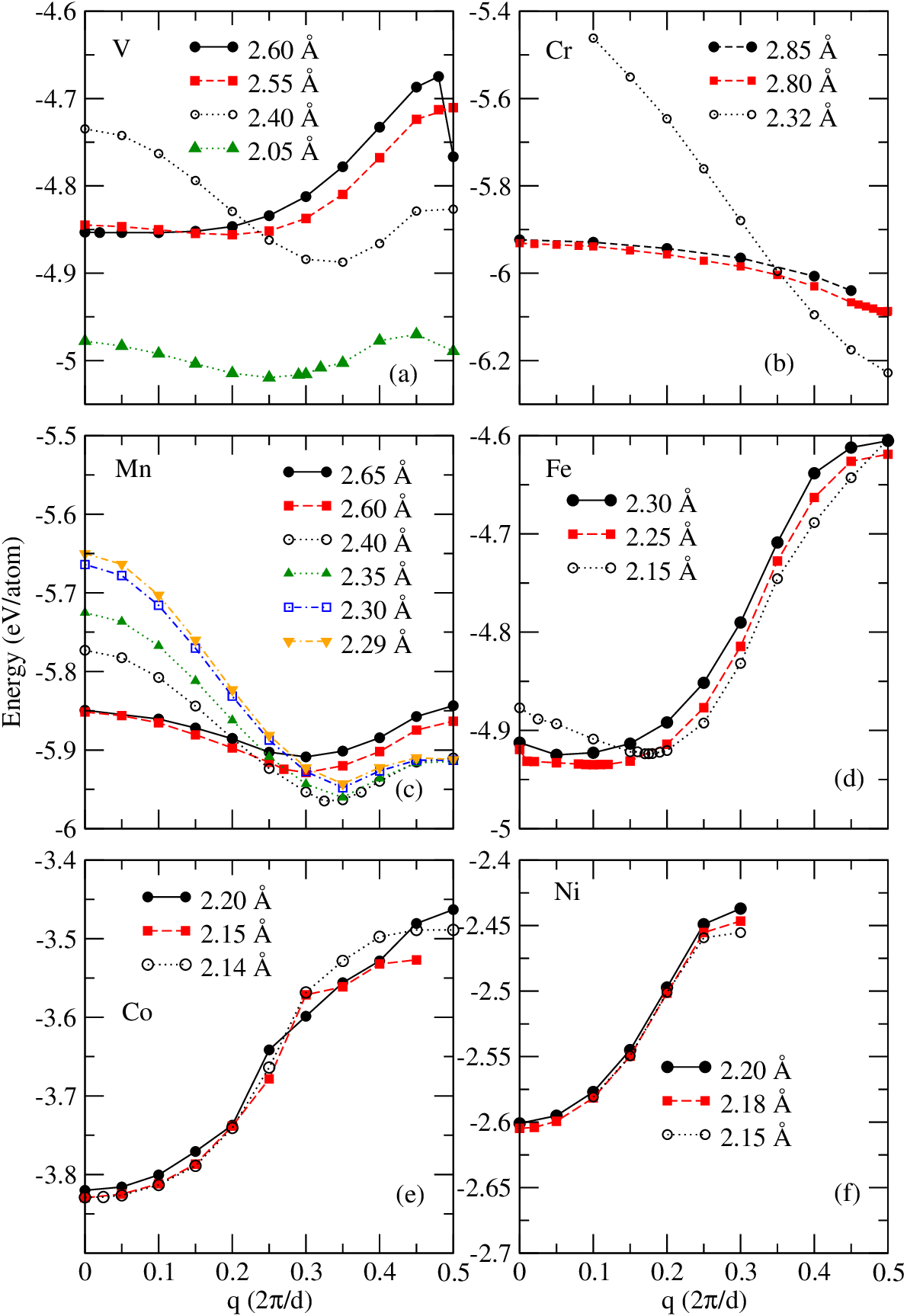}\\
\caption{ (color online) Total energy $E(q)$ versus spin-spiral
wave vector $q$ of the 3$d$ transition metal chains at several different bond lengths $d$.}
\label{spiralQ}
\end{figure}

The calculated total energies [$E(q,\theta)$] as a function of the spin-spiral propagation vector $q$ 
of the 3$d$ transition metal chains at several different bond lengths $d$ are
plotted in Fig. \ref{spiralQ}. Since we consider here the transverse spin-spiral waves only,
the angle between the chain axis (i.e., $z$-axis) and the magnetization direction $\theta = \pi/2$,
and hence we simply write $E(q,\theta = \pi/2) = E(q)$.
The spin-spiral structure at wavevector
vector $q =0$  corresponds to the collinear ferromagnetic (FM) state,
whilst the state at $q = 0.5$ (2$\pi$/$d$) corresponds to  the antiferromagnetic (AF) state.
Therefore, as shown in Fig. \ref{spiralQ}, at $q =0$ the lowest
total energy state of the Cr (Mn) chain occurs at 2.80 (2.60) \AA, but it appears at 2.32 (2.29) \AA$ $
 at $q=0.5$ (2$\pi$/$d$), being in good agreement with our previous collinear magnetic calculations\cite{Tung}.
Interestingly, Fig. \ref{spiralQ} shows that both the FM and AF states
in the V, Mn and Fe chains become unstable against formation of a spin-spiral structure.
Furthermore, the lowest total energy of the spin-spiral state occurs at the bond length that is generally
different from that of the collinear magnetic states. For example, the ground state of the Mn
chain is the spin-spiral state with the equilibrium bond length of 2.40 \AA, instead
of 2.60 \AA$ $ (the FM state) and 2.29 \AA$ $ (the AF state)\cite{Tung}.

Nonetheless, there is no stable spin-spiral state in the Cr, Co and Ni chains.
In the Cr chain, therefore, the AF state remains the stable state. In the Co and Ni chains,
the FM state still has the lowest total energy (see Fig. \ref{spiralQ}).
In fact, we could not even obtain a spin-spiral solution for the Ni chain at the wave vector $q$ being larger than
0.3 (2$\pi$/$d$). This is because, as shown in Fig. \ref{spiralQ}, an increase in the number of valence
electrons leads to an increased stabilization of the FM state whilst a decrease in the number of valence
electrons tends to stabilize the AF state.
This observation is further corroborated by the fact that in the Cr chain of bond length $d = 2.32$ \AA,
the total energy decreases steeply as the spin spiral wave vector increases (Fig. \ref{spiralQ}b).
For comparison, we notice that in previous GGA calculations\cite{Guo00}, the FM state could not be stabilized in
bulk Cr metal, whilst the magnetization energy of the AF state is rather small ($\sim$0.016 eV/atom).

The calculated equilibrium bond length ($d$), ground state wave vector ($q$) and
total energy $[E(q)]$ as well as spin magnetic moment per unit cell ($m_s$) at $q=0$ of all
the 3$d$ transition metal chains considered here are listed in Table \ref{table1}.
We notice that the spin magnetic moment of the spin-spiral state of the V chain is significantly reduced,
in comparision with the FM state (see Ref. \onlinecite{Tung}), but is close to that of the AF state.
Interestingly, Table \ref{table1} shows that the stable spin-spiral wave length $\lambda$ of the Fe chain is nearly exactly
10 bond lengths (or lattice constant),
whilst, in constrast, that for the V and Mn chains is much shorter, being 4 and 3 bond lengths, respectively.
However, the spin-spiral wave length $\lambda$ can depend on the bond length, and this dependence
is especially pronounced for the V chain, as demonstrated in Fig. \ref{spiralQ}a.
When the V chain is stretched to the bond length of 2.40 \AA,
the spiral propagation vector $q$ becomes $\sim$0.35 (2$\pi$/$d$), but when it is further stretched to $d = 2.55$ \AA,
$q$ is reduced to $\sim$0.20 (2$\pi$/$d$). A similar behavior of the spin-spiral wave vector can be found for the Fe chain,
as shown Fig. \ref{spiralQ}d.

\begin{figure}
\includegraphics[width=8cm]{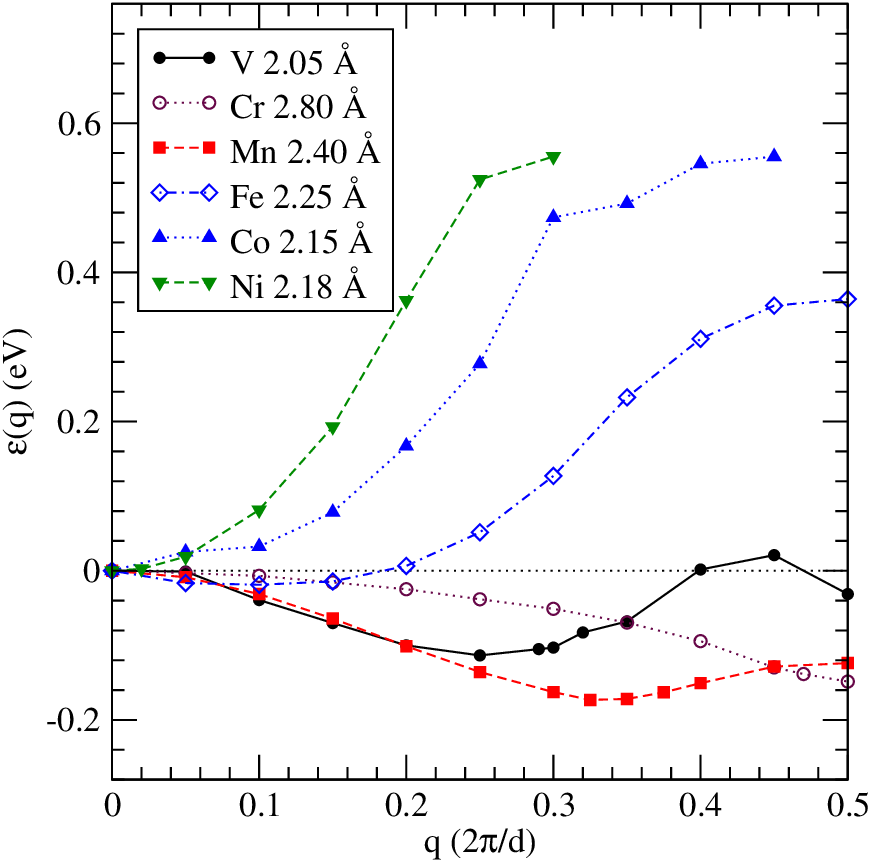}\\
\caption{ (color online) Calculated spin-wave energy spectra $\varepsilon (q)$ $[$i.e., magnon dispersion relations
$\hbar \omega (q)$$]$ of the 3$d$ transition metal atomic chains at the ground state bond length.}
\label{3dJ}
\end{figure}

The energy of a spin-wave excitation (i.e., the magnon dispersion relation) is
given as the derivative of the total energy of the spin-spiral state with respect to
the magnon number \cite{Rosengaard,Niu99}
\begin{equation}\label{spinwavew}
  \varepsilon(q) = \hbar\omega(q)=2\mu_B\frac{\Delta E({\bf q},\theta)}{\Delta M} = \frac{4\mu_B}{m_{s0}}[E(q)-E(0)]
\end{equation}
where $\Delta E(q,\theta) = E(q,\theta) - E(0,\theta) = E(q) - E(0)$ is the energy of a spin spiral of wave vector ${\bf q}$
relative to the ferromagnetic state ($q=0$),
$\Delta M$ is the decrease of the magnetization per site
projected onto the $z$-axis, and $m_{s0}$ is the spin magnetic moment per site at $q = 0$.
The calculated magnon dispersion relations for the V, Cr, Mn, Fe, Co and Ni chains
at the minimal energy bond length are plotted in Fig. \ref{3dJ}.

{\it Ab initio} calculations of the excitation energy of the spin-spiral wave
along some high-symmetry lines in the Brillouin zone in bulk bcc Cr\cite{Shallcross},
fcc Mn\cite{Shallcross}, bcc Fe\cite{Shallcross,Pajda, Rosengaard,
Halilov}, fcc Co\cite{Shallcross, Pajda,Rosengaard,Halilov},
and fcc Ni\cite{Shallcross,Pajda, Rosengaard,Halilov} metals have been reported before,
and the stable spin-spiral structures were found in bcc Cr and fcc Mn. The stable spin-spiral wave with
a wave length of $\sim$7 lattice constants was also found in a freestanding bcc Fe(110)
monolayer with lattice constant of 3.16 \AA$ $ in two previous {\it ab initio} studies\cite{Mizuno, Nakamura}.
Here we predict the existance of the stable spin-spiral structures in 1D freestanding transition metal
(V, Mn and Fe) atomic chains for the first time.

\section{Exchange interactions}
\begin{table}
\caption{Calculated exchange interaction parameters ($J_{0j}$) (meV) between two $j^{th}$ near neigbors ($j = 1, 2, 3, 4, 5$)
in the V, Cr, Mn, Fe, Co and Ni atomic chains.}
\begin{ruledtabular}
\begin{tabular}{cccccc}
    & $J_{01}$  & $J_{02}$  & $J_{03}$  &  $J_{04}$  &  $J_{05}$  \\ \hline
  V  &   4.2        &  -22.8       &   -0.2      &   2.6      &  -2.4       \\
  Cr & -65.0        &   17.6       &   -7.2      &   4.0      &  -2.6       \\
  Mn & -78.4        &  -43.2       &    9.8      &  -3.0      &   1.2       \\
  Fe & 158.2        &  -57.8       &    4.4      &   2.6      &  -3.2       \\
  Co & 156.4        &   13.0       &  -22.8      &  16.6      & -11.8       \\
  Ni & 109.0        &    6.6       &   20.0     &  -24.6      &   7.4       \\
%
\end{tabular}
\end{ruledtabular}
\label{tableJ}
\end{table}

To a rather good approximation, we can map a metallic magnet onto an effective Heisenberg Hamiltonian
with classical spins\cite{Halilov,Pajda}
\begin{equation}\label{heisenberg}
    H_{eff} = -\frac{1}{2}\sum_{i, j}J_{ij}{\bf \sigma}_i\cdot{\bf \sigma}_j
\end{equation}
where $J_{ij}$ is an exchange interaction parameter between atomic site $i$ and site $j$,
and ${\bf \sigma_i}$ (${\bf \sigma_j}$) is the unit vector representing the direction of
the local magnetic moment at site $i$ ($j$). 
In the frozen magnon approach, the exchange interaction parameters $J_{ij}$ are related to
the magnon excitation energy $\varepsilon ({\bf q})$ by a Fourier transformation
\begin{equation}\label{Jexchange}
    J_{0j}=\frac{1}{N_{\bf q}}\sum_{\bf q}e^{-i{\bf q}\cdot{\bf R}}J({\bf q})
\end{equation}
where $N_{\bf q}$ is the number of ${\bf q}$ points in the Brillouin zone included in the summation and
\begin{equation}\label{frozenE}
   \varepsilon({\bf q}) = \frac{4\mu_B}{m_{s0}}[E({\bf q}) - E(0)]=-\frac{2\mu_B}{m_{s0}}sin(\theta)^2J({\bf q}) = -\frac{2\mu_B}{m_{s0}}J({\bf q}).
\end{equation}
Here, $\theta$ is fixed to $\pi/2$ for all the spin-spiral states and $J({\bf q})$ is the Fourier transform of the exchange parameters.
We therefore evaluate the exchange interactions in the 3$d$ transition metal chains via Eqs. \ref{Jexchange} and \ref{frozenE}
by using the calculated magnon dispersion relations, as shown in Fig. \ref{3dJ}.

\begin{figure}
\includegraphics[width=8cm]{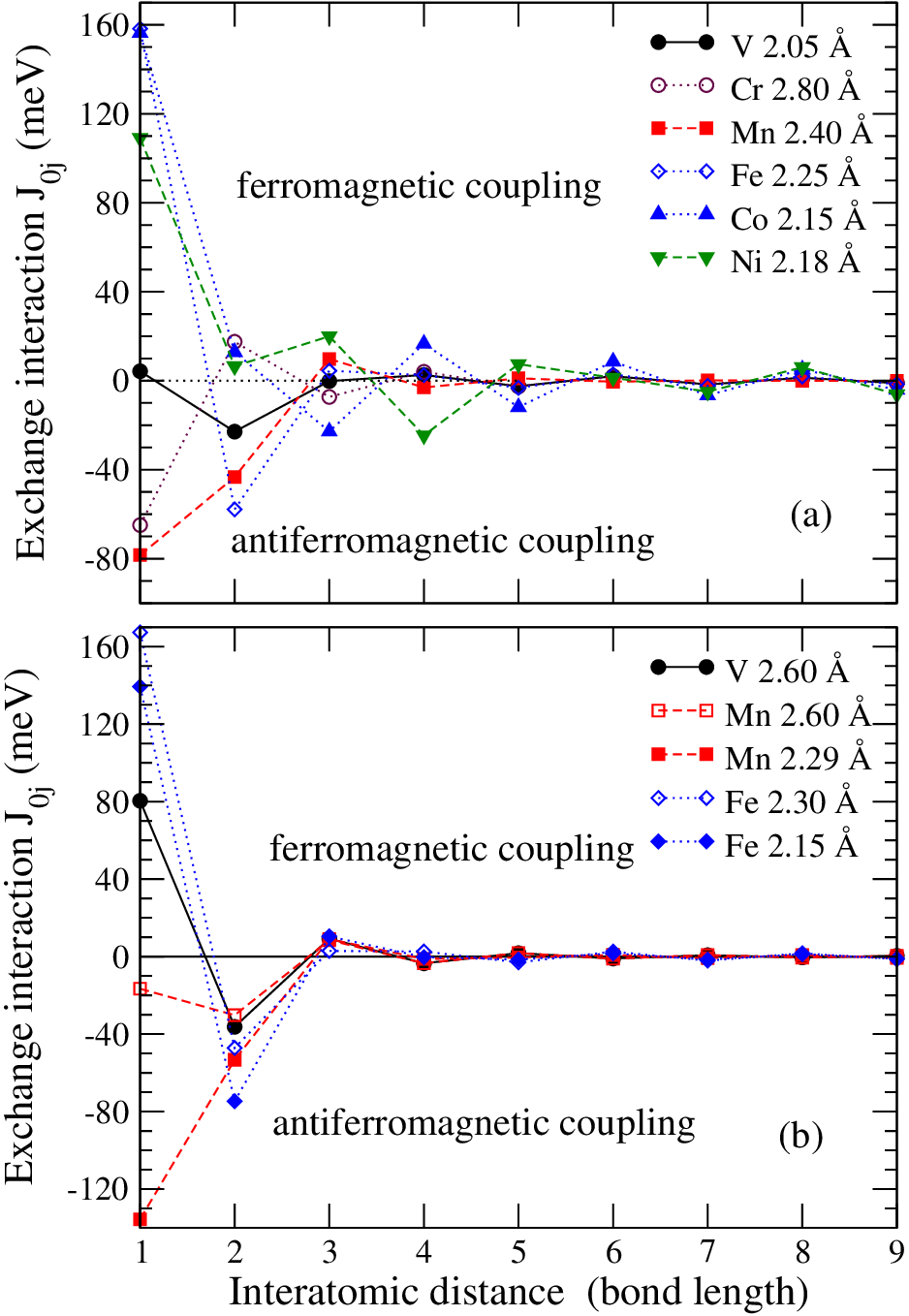}\\
\caption{ (color online) Calculated exchange interaction parameters $J_{0j}$ versus interatomic 
distance for the 3$d$ transition metal chains with different bond lengths as labeled.
In (a), the atomic chains are in their respective ground state bond lengths, and in (b), 
the atomic chains are in either stretched or compressed bond lengths.}
\label{exchangeJ}
\end{figure}

The obtained exchange interaction parameters as a function of the interatomic distance are plotted in Fig. \ref{exchangeJ} and
also listed in Table \ref{tableJ}.
In the minimum energy bond lengths, as shown in Fig. \ref{exchangeJ}a, the magnetic coupling
between two first nearest neighbors in the V, Fe, Co and Ni chains is ferromagnetic ($J_{01} > 0$),
whilst it is antiferromagnetic ($J_{01} < 0$) in the Cr and Mn chains.
In the Co and Ni chains, the magnetic coupling between the second nearest neighbors remain
ferromagnetic ($J_{02} > 0$) and this explains why the ground state of these chains is ferromagnetic.
In contrast, the magnetic coupling between the second nearest neighbors in the V, Mn and Fe chains is
antiferromagnetic ($J_{02} < 0$), i.e., the exchange interactions in these chains would be frustrated.
As a result, noncollinear spin-spiral states in these chains may become energetically more favorable
than either collinear ferromagnetic or antiferromagnetic state. Let us take the Mn chain as an example,
and consider the atomic spin at the origin.
This spin tends to couple antiferromagnetically both with its two nearest neighbors and also with its two second near neighbors
because the first and second near neighbor exchange parameters $J_{01}$ and $J_{02}$ are negative (see Fig. \ref{exchangeJ}a and
Table \ref{tableJ}).
However, this would make its two nearest neighbor spins "frustrated" because they would have to couple
ferromagnetically with one nearest neighbor on one side and antiferromagnetically with the other nearest neighbor on the
opposite side. This frustrated magnetic coupling therefore would energetically favor a spin-spiral state.
In fact, according to the mean field theory for an 1-D classical Heisenberg spin chain with
the negligibly small magnetic coupling between third near neighbors and beyond (see, e.g., Ref. \onlinecite{James}),
in the frustrated magnetic coupling situation ($J_{02} < 0$), the system would be ferromagnetic
($q = 0$) if $J_{01} > 4|J_{02}|$ and would be antiferromagnetic ($q = \pi/2$) if $J_{01} < -4|J_{02}|$.
Table \ref{tableJ} shows clearly that the condition $J_{01} > 4|J_{02}|$ is fulfilled for the Co and Ni chains,
giving rise to the ferromagnetic ground state, and  that the condition $J_{01} < -4|J_{02}|$ is met for the Cr chain,
resulting in the antiferromagnetic ground state.
Interestingly, a stable spin-spiral structure with the spiral propagation vector $q$ given by
$\cos(qd)=-J_1/4J_2$ would occur if $J_{02} < 0$ and $4J_{02} < J_{01} < 4|J_{02}|$.\cite{James}
Using the exchange coupling parameters listed in Table \ref{tableJ},
we would obtain the spiral propagation vector $q = 0.24, 0.32$ and $0.13$ ($2\pi/d$), respectively,
for the V, Mn and Fe chains. These estimated $q$ values agree very well with that obtained by
the fully selfconsistent total energy calculations (Table \ref{table1}).

Figure 1 shows that the stability of spin-spiral state in the V, Mn and Fe chains depends 
pronouncedly on bond length $d$. Therefore, we also calculate the exchange interaction 
parameters for these chains with several other bond lengths and 
the results are plotted in Fig. 3(b). The equilibrium bond length for the FM V and Mn chains is 
2.60 \AA. The equilibrium bond lengths for the AF Mn and Fe chains are 2.29\AA$ $ and 2.15\AA,
respectively. It is clear from Fig. 3 that the calculated exchange interaction parameters
can be sensitive to the bond length. For example, the magnitude of the nearest neighbor 
exchange interaction $J_{01}$ in the V atomic chain is dramatically increased from 2.1 meV 
to 40 meV as the bondlength is increased from 2.05 \AA$ $ to 2.60 \AA. A similar behavior is found
for the Fe chain (Fig. 3). Interestingly, in contrast, the magnitude of the nearest neighbor
exchange interaction $J_{01}$ in the Mn chain is significantly reduced as the bond length is 
increased. Nevertheless, the exchange interaction parameters for the second near-neighbor and
beyond are less affected by the bond length (see Fig. 3).


{\it Ab initio} evaluation of the exchange interaction parameters in bulk ferromagnets Fe, Co and Ni have
been reported many times before.
For example, the nearest neighbor exchange interaction $J_{01}$ in bcc Fe was determined to be
39.0 meV\cite{Pajda} and 57.3 meV\cite{Wang}, respectively. The nearest neighbor exchange interaction
$J_{01}$ in fcc Co was estimated to be 29.5 meV\cite{Pajda}. In bulk fcc Ni, the $J_{01}$ was calculated to
be 5.6 meV\cite{Pajda}. In the freestanding square Fe and Co monolayers with the Cu(001) lattice constant,
the nearest neighbor exchange interaction $J_{01}$ is 92.5 and 77.6 meV\cite{Pajda2}, respectively.
These indicate that in 2D systems, in general, the exchange interactions are significantly
enhanced as compared with their bulk counterparts, mainly because of reduced coordination numbers.
Table II show that the nearest neighbor exchange parameters $J_{01}$ in the Fe, Co and Ni chains,
are much larger than in their bulk counterparts. This may be attributed, at least partially, to
the fact that the nearest bond length in the atomic chains is significantly shorter than in their
bulk counterparts\cite{Tung}. Our calculated $J_{01}$ values in the Fe and Co chains
are also significantly larger than the corresponding $J_{01}$ values in the freestanding Fe and Co monolayers\cite{Pajda2}.
Interestingly, in the Fe atomic chains deposited on the Cu(117) surface,
the effective exchange interaction parameter $J_{eff}$ was determined to be
around 136 meV\cite{Spisak}, being comparable with the corresponding $J_{01}$ value of the
freestanding Fe chain listed in Table II.

\section{Spin-wave stiffness and critical temperature}

The calculated energy dispersion relations of the spin-spiral waves $\varepsilon(q)$
for the 3$d$ transition metal chains at the ground state bond length are displayed in Fig. \ref{3dJ}.
In the range of small $q$, $\varepsilon(q) = Dq^2$, where the spin-wave stiffness constant $D$
relates the spin-wave energy $\hbar \omega (q)$ to the wave vector $q$ in the long wavelength limit.
The spin-wave stiffness constant $D$ of an atomic chain can be estimated by fitting an even order polynormial to
the corresponding spin-wave spectrum shown in Fig. \ref{3dJ}.
The spin-wave stiffness constant $D$ obtained in this way for the 3$d$ metal chains are listed in Table \ref{tableD}.
A negative value of $D$ means that the FM state is not stable against a spin-spiral wave excitation.
Table \ref{tableD} shows that the spin-wave stiffness constant $D$ in the V, Cr, Mn and Fe chains is negative.
Only in the Co and Ni chains, the $D$ is positive.

In principle, one can also calculate the spin-wave stiffness constant $D$ via\cite{Rosengaard}
  \begin{equation}\label{spinwaveD}
    D =\frac{2\mu_B}{m_{s0}}\frac{d^2 E(q)}{dq^2}=\frac{\mu_B}{3m_{s0}}\sum_{j}J_{0j}R_{0j}^2
\end{equation}
where $J_{0j}$ are the exchange interaction parameters and $R_{0j} = |{\bf R}_0-{\bf R}_j|$ is the
distance between site 0 and site $j$. In practice, Eq. \ref{spinwaveD} cannot be used directly to obtain
reliable values for the spin-wave stiffness constant, because the numerical uncertainties at the long
distances are amplified by the factor $R_{0j}^2$.\cite{Pajda} Here we use this expression to understand the calculated
$D$s listed in Table \ref{tableD}. For example, the magnitude of the spin-stiffness constant $D$ of the V chain
is much larger than that of the Cr chain because the V chain has a much smaller spin magnetic moment
(see Table I) and also a negative second near neighbor antiferromagnetic exchange parameter (see Table \ref{exchangeJ}).
Furthermore, even though $J_{01}$ is positive in the
V and Fe chains, the $D$ is negative, because the V and Fe chains have $J_{02} < 0$ and $J_{01} < 4|J_{02}|$.

\begin{table}
\caption{Calculated spin wave stiffness constant $D$ (meV\AA$^2$) and magnetic transition
temperature $T_C$ of the 3$d$ metal chains.
Also listed are the $D$s and $T_C$s for the 3D and 2D metal systems from previous {\it ab initio} calculations and
experimental measurements, for comparision.}
\begin{ruledtabular}
\begin{tabular}{ccccccc}
          & \multicolumn{3}{c}{stiffness $D$}& \multicolumn{3}{c}{$T_C$ (K)} \\
          & 3D & 2D & 1D &  3D & 2D & 1D \\ \hline
  V  & - & - & -424  & - & - & 94 \\
  Cr & - & - & -106  & 311\footnotemark[7] & - &414 \\ 
  Mn & - & - & -504  & - & - & 274 \\
  Fe & 250\footnotemark[1], 330\footnotemark[2] & 164\footnotemark[3]  & -78 & 1414\footnotemark[1], 1043\footnotemark[4] & 1265\footnotemark[3] & 410 \\
  Co & 663\footnotemark[1], 510\footnotemark[2] & 570\footnotemark[3], 427\footnotemark[5]  & 616  & 1645\footnotemark[1], 1388\footnotemark[4] & 1300\footnotemark[3] & 606 \\
  Ni & 756\footnotemark[1], 555\footnotemark[6], & - &656  & 397\footnotemark[1], 627\footnotemark[4]  &  & 458
%
\end{tabular}
\footnotetext[1]{Theoretical calculations (Ref. \cite{Pajda}).}
\footnotetext[2]{Neutron-scattering measurement extrapolated to 0 K (Ref. \cite{Pauthenet}).}
\footnotetext[3]{Theoretical calculations (Ref. \cite{Pajda2}).}
\footnotetext[4]{Experimental measurements (Ref. \cite{Ashcroft}).}
\footnotetext[5]{Brillouin light scattering measurement (Ref. \cite{Akira}).}
\footnotetext[6]{Neutron-scattering measurement (Ref. \cite{Mook}).}
\footnotetext[7]{Neutron-scattering measurement (Ref. \cite{Fawcett}).}
\end{ruledtabular}
\label{tableD}
\end{table}

For comparision, the spin-wave stiffness constants $D$s for the three-dimensional (3D) and two-dimensional (2D)
Fe, Co and Ni systems from previous {\it ab initio} calculations and experimental measurements, are
also listed in Table \ref{tableD}. It is clear from Table \ref{tableD} that the spin-wave stiffness constant tends to become smaller
as the dimensionality of the system gets reduced. This may be expected (see Eq. \ref{spinwaveD})
because the number of near neighbors becomes
smaller as the dimensionality of the system gets reduced. Among the three bulk 3$d$ elemental ferromagnets, Fe has
the smallest spin-wave stiffness, and, interestingly, the stiffness of Fe becomes negative when its dimensionality
is decreased to one (Table \ref{tableD}).

Within the mean-field (MF) approximation, the critical temperature ($T_C$) of the magnetic phase transition
for the effective Heisenberg Hamiltonian can be estimated via the approximate expression\cite{CSWang,Jensen}
\begin{equation}\label{transMF}
k_BT_C^{MF}=\frac{1}{3}J({\bf q})
\end{equation}
where ${\bf q}$ is the spin-spiral wavevector and $J({\bf q})$ is the Fourier transform of
the interatomic exchange parameters
\begin{equation}\label{exJq}
J({\bf q}) = \sum_{j}{J_{0j}e^{i{\bf q}\cdot {\bf R}_{0j}}}.
\end{equation}
In the ferromagnetic case (${\bf q} = 0$)\cite{CSWang},
\begin{equation}\label{CurieFM}
 k_BT_C^{MF}= \frac{m_{s0}}{6\mu_B N_{\bf q}}\sum_{{\bf q}}{\varepsilon(\bf{q})}.
\end{equation}
Using the calculated exchange interaction parameters $J_{0j}$ (Table \ref{tableD} and Fig. \ref{exchangeJ}a),
and Eqs. \ref{transMF} and \ref{exJq}, we estimate the transition temperatures $T_C^{MF}$ for the 3$d$ transition metal chains,
as listed in Table \ref{tableD}. Encouragingly, the ferromagnetic transition temperatures evaluated by using Eq. \ref{CurieFM}
for the Co and Ni chains are 622 and 444 K, respectively, being in good agreement with that obtained by
using Eqs. \ref{transMF} and \ref{exJq} (Table \ref{tableD}).
Table \ref{tableD} indicates that the critical temperature
for the 3$d$ atomic linear chains considered varies from several tens to a few hundreds of Kelvins.
The critical temperatures for the 3$d$ atomic chains are much smaller than the corresponding ones
for the bulk metals and their monolayers (Table \ref{tableD}). This is because of the much reduced coordination
numbers in these 1D systems, although the exchange interaction parameters in the atomic chains are generally
larger than their counterparts in the bulks and monolayers.

The more accurate expression for the critical temperature within the random phase approximation (RPA) also
exists\cite{CSWang,Pajda2}
\begin{equation}\label{CurieRPA}
\frac{1}{k_BT_C^{RPA}}=\frac{6\mu_B}{M}\frac{1}{N_q}\sum_{q}\frac{1}{\varepsilon(\bf{q})+\Delta}
\end{equation}
where $\Delta$ is the magnetic anisotropy energy.
The RPA expression generally gives the critical temperatures for bulk ferromagnets
Fe, Co and Ni being in better agreement with experiments than the MF expression\cite{Pajda}.
Furthermore, the critical temperatures with the MF approximation are usually significantly higher than
that from the RPA calculations.\cite{Pajda2,Pajda} Using our calculated spin-wave dispersion relations (Fig. 2)
and also the $\Delta$ values from Ref. \onlinecite{Tung}, we obtain $T_C^{RPA} = 230$ K for the Co chain
and $T_C^{RPA} = 229$ for the Ni Chain. 
Therefore, the estimated critical temperatures listed in Table \ref{tableD}
should be considered only as the upper limits.
Finally, we note that in 1-D isotropic Heisenberg model
with finite-range exchange interactions, there is no spontaneous magnetization at any nonzero
temperature because fluctuations become important.\cite{Mermin}
Nonetheless, this discouraging conclusion has to be revised in the
presence of a magnetic anisotropy and long range interactions.
Indeed, ferromagnetism in 1-D monoatomic Co metal chain on a Pt substrate has been recently
reported.\cite{Gam02} A detailed discussion on possible finite temperature spontaneous magnetization
in 1-D systems has been given in Ref. \onlinecite{Cur05}.




\section{Electronic band structure}
\begin{figure}
\includegraphics[width=8cm]{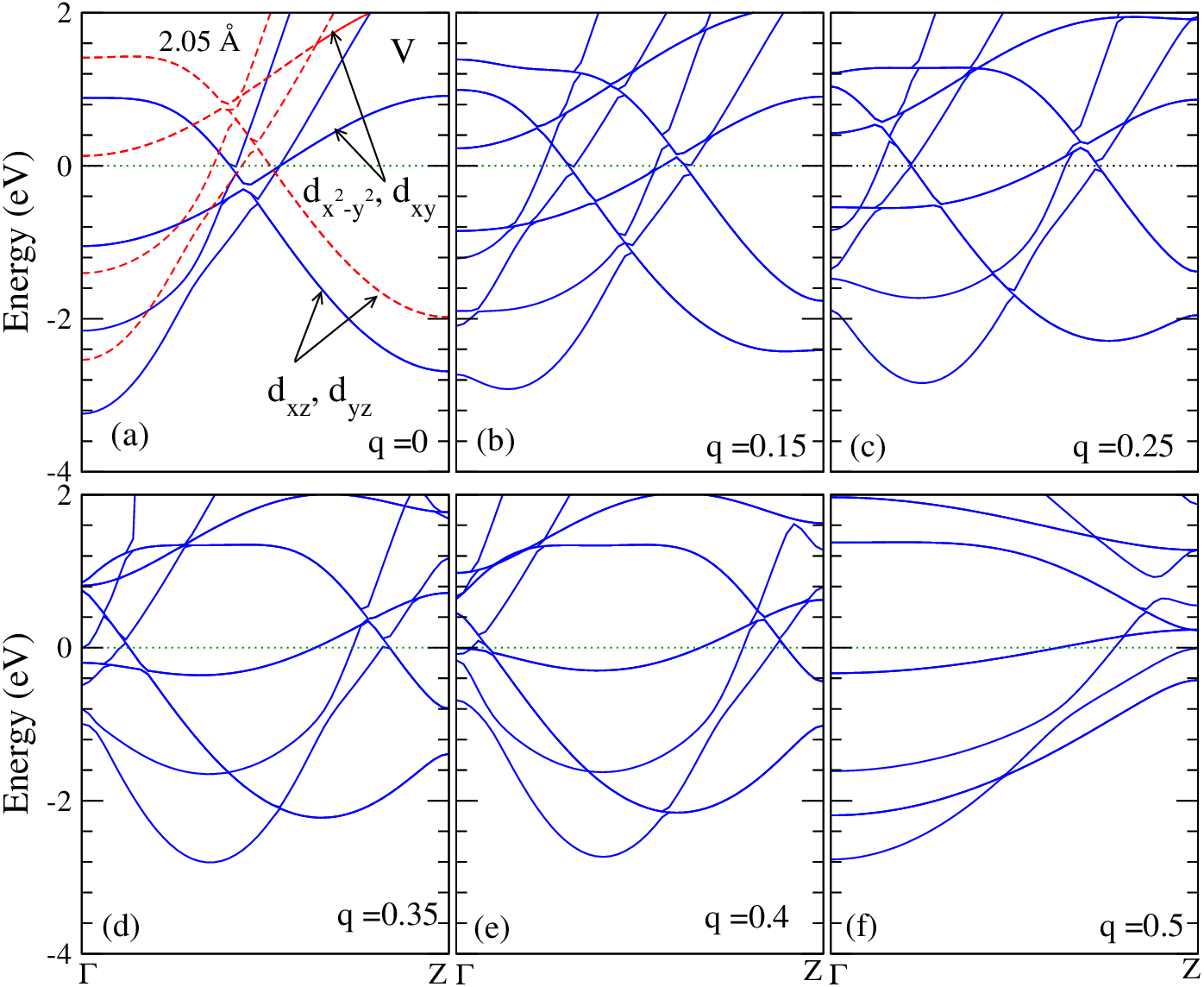}\\
\caption{(color online) Band structures of the V chain in different spiral propagation vector $q$ along $\Gamma$ to $Z$ direction.
Figure (a), $q=0$ ($2\pi/d$), that indicates a spin-polarized ferromagnetic band structure whilst figure (f), $q=0.5$ ($2\pi/d$), indicates an AF
band structure.}
\label{VBand}
\end{figure}

\begin{figure}
\includegraphics[width=8cm]{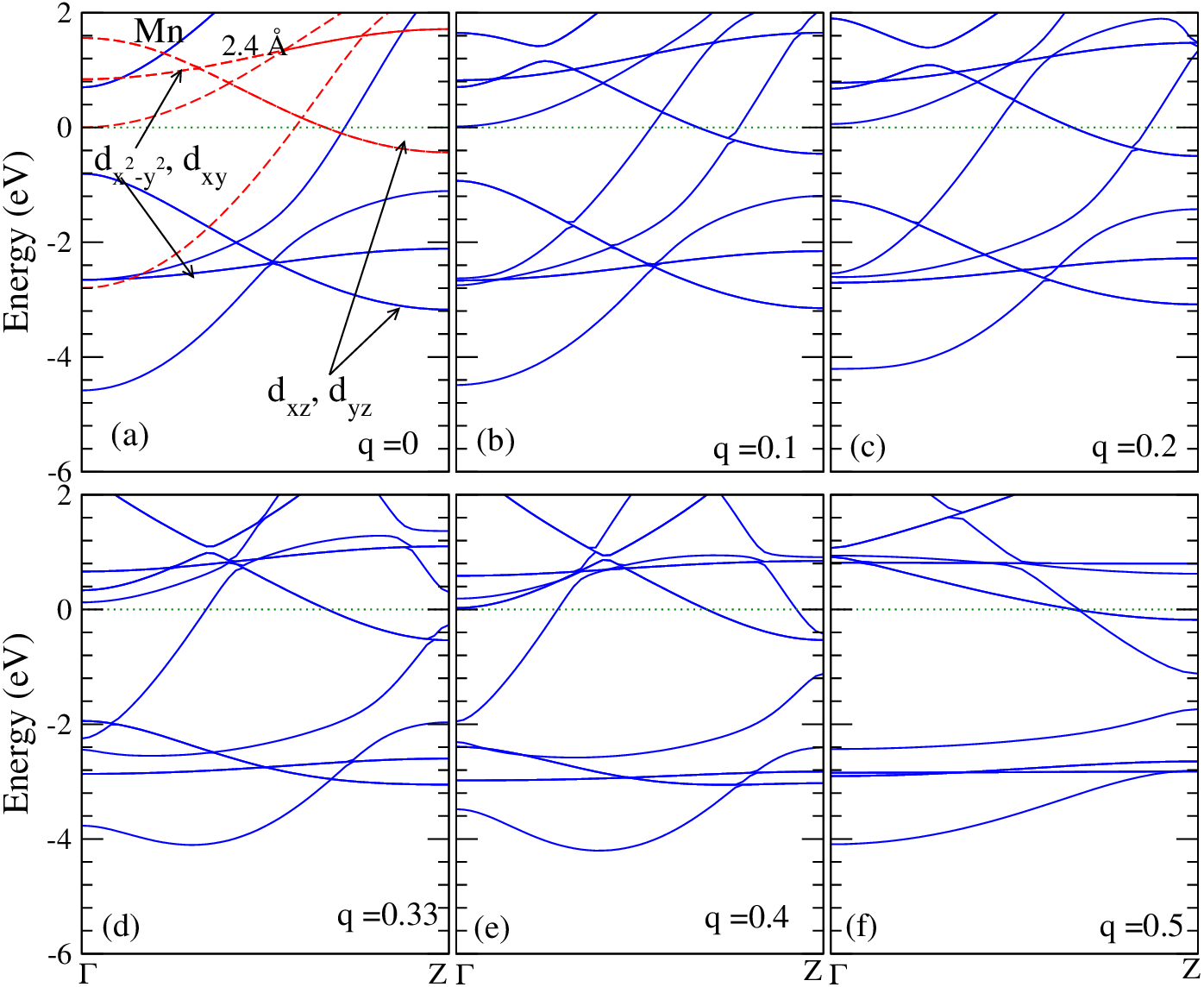}\\
\caption{(color online) Band structures of Mn in different spiral propagation vector $q$ along $\Gamma$ to $Z$ direction.
Figure (a), $q=0$ ($2\pi/d$), that indicates a spin-polarized ferromagnetic band structure whilst figure (f), $q=0.5$ ($2\pi/d$), indicates an AF
band structure.}
\label{MnBand}
\end{figure}

In order to study how the spin-spiral structure affect the electronic band structure\cite{Sandratskii}
 and also to
help getting further insight into the spin-spiral instability at the microscopic level,\cite{Kor96,Lizarraga2}
the electronic band structures of
the V and Mn chains at several spin-spiral wave vectors $q$ are displayed in
Fig. \ref{VBand} and Fig. \ref{MnBand}, respectively.
The ferromagnetic ($q = 0$) band structures are presented in Fig. \ref{VBand} and Fig. \ref{MnBand}a. Because of the linear
chain symmetry, the bands may be grouped into three sets, namely, the nondegenerate $s$- and $d_{z^2}$-dominant bands,
doubly degenerate ($d_{xz}$, $d_{yz}$) bands, and ($d_{x^2-y^2}$ ,$d_{xy}$) dominant bands,
see Figs. \ref{VBand}a and \ref{MnBand}a.
The ($d_{x^2-y^2}$, $d_{xy}$) bands are narrow because the $d_{x^2-y^2}$, and $d_{xy}$ orbitals are perpendicular
to the chain, thus forming weak $\gamma$ bonds. The $d_{xz}$, and $d_{yz}$bands, on the other hand, are more
dispersive due to the stronger overlap of the $d_{xz}$ and $d_{yz}$ orbitals along the chain, which gives rise to
the $\pi$ bonds. The $s$- and $d_{z^2}$-dominant bands are most dispersive since these orbitals form strong $\sigma$
bonds along the chain.

Two main changes could appear in a ferromagnetic band structure when a spin-spiral wave is introduced,
namely, the lifting of the accidental degeneracy at the cross-point of spin-up and spin-down bands
and the "repulsion" of opposite-spin bands.\cite{Sandratskii}
These changes can be clearly seen in Figs. \ref{VBand}b and \ref{MnBand}b. For example, in Fig. \ref{VBand}b,
the spin-up and spin-down $d_{xz}$ and $d_{yz}$ bands clearly move away from each other.
Of course, these changes due to the noncollinear spin-spiral wave become more pronounced as $q$ increases
(Figs. \ref{VBand} and \ref{MnBand}). In the V chain, the repulsion of opposite-spin bands appear to
lower the spin-up $d_{xz}$-$d_{yz}$ band considerably while in the mean time raise the spin-down
$d_{xz}$ and $d_{yz}$ band significantly (see, e.g., Figs. \ref{VBand}b and c).
This could be the reason why the spin-spiral structure
is energetically favored over the ferromagnetic state.
In the Mn chain, on the other hand, as the wave vector $q$ becomes nonzero, the fully occupied
spin-up $d_{xz}$-$d_{yz}$ band near the $\Gamma$ point is pushed down substantially
(see Fig. \ref{MnBand}).
This change could lead to a lowering
of the total energy, thereby stabilizing the spin-spiral structure in the Mn chain.
Figs. \ref{VBand} and \ref{MnBand} also show, in contrast, that as the spiral wave vector $q$ becomes nonzero,
the dispersive $s$- and $d_{z^2}$-dominant valence bands near the $\Gamma$ point are pushed up in energy,
and these shifts are more or less proportional to the modulus of spin spiral vector ${\bf q}$.
This upward movement of the  $s$- and $d_{z^2}$-dominant bands might raise the total band energy.
Therefore, the final equilibrium spiral wave vector $q$ would be determined by
a trade-off of these two contrasting changes in the electronic band structure of the atomic chain.


\section{Conclusions}
We have calculated the total energy of the transverse spin-spiral wave as a function of the wave vector
for all 3$d$ transition metal atomic chains within  {\it ab initio} density functional theory with generalized
gradient approximation. As a result, we predict that at the equilibrium bond length, the V, Mn, and Fe
chains have a stable spin spiral structure.
Furthermore, all the exchange interaction parameters of the 3$d$ transition metal chains are evaluated
by using the calculated energy dispersion relations of the spin-spiral waves. Interestingly, we find that
the magnetic couplings in the V, Mn and Cr chains are frustrated (i.e., the second near neighbor exchange
interaction is antiferromagnetic), and this leads to the formation of the stable spin-spiral structure
in these chains. The spin-wave stiffness constant of these 3$d$ is also evaluated and compared with
its counterpart in bulk and monolayer systems.
We have also estimated the upper limit of the possible magnetic phase transition temperature
in these atomic chains within the mean field approximation.
The electronic band structure of the spin-spiral structures have also been calculated.
We hope that our findings of the stable spin-spiral structure and frustrated magnetic interaction in
the 3$d$ transition metal chains would stimulate further theoretical and experimental research in
this field. Indeed, after learning our {\it ab initio} results, Sandvik recently studied a spin-1/2
Heisenberg chain with both frustration and long-range interactions by exact diagonalization.\cite{San10}
He found a first-order transition between a Neel state and a valence-bond-solid with coexisting
critical $k = \pi/2$ spin correlations.\cite{San10}


\section*{Acknowledgments}
The authors thank A. W. Sandvik and Z. R. Xiao for stimulating discussions.
The authors acknowledge support from the National Science Council and the NCTS of Taiwan. They also thank the
National Center for High-performance Computing of Taiwan and the NTU Computer and Information Networking Center
for providing CPU time.


\end{document}